\documentclass[aps,prl,twocolumn]{revtex4}
\usepackage{amsmath}
\usepackage{amssymb}

\newcommand{\ket}[1]{\left| {#1} \right\rangle}
\newcommand{\bra}[1]{\left\langle {#1}\right |}
\newcommand{\iprod}[2]{\left\langle {#1}|{#2}\right\rangle}
\newcommand{\proj}[1]{\left| {#1} \right\rangle \left\langle {#1}\right |}

\newcommand{\Pa}[1]{\mathcal{P}_{{#1}}}

\def\p{\ket{\psi}}
\def\S{S(\psi)}

\def\be{\begin{equation}}
\def\ee{\end{equation}}
\def\bea{\begin{eqnarray}}
\def\eea{\end{eqnarray}}

\def\>{\rangle}
\def\<{\langle}

\begin{document}

\title{Strong Super-additivity of the entanglement of formation for pure stabilizer states}

\author{David Fattal}
\affiliation{Hewlett-Packard Laboratories, 1501 Page Mill Road, Palo
Alto, CA 94304}
\author{Keiji Matsumoto}
\affiliation{NII, Tokyo...}

\date{\today}

\begin{abstract}
We prove the strong super-additivity of the entanglement of
formation for stabilizer pure states, and the set of mixed states
which minimize their average entropy of entanglement as a mixture of
stabilizer pure states sharing the same stabilizer group up to
phases. The implications of the result on the additivity of the
Holevo capacity of a quantum channel transmitting stabilizer states
with Pauli noise is discussed.
\end{abstract}

\pacs{}
\maketitle


%
%
%

Among the open problems of quantum information theory, the
additivity of the Holevo capacity of a quantum channel is an
important one. The Holevo capacity is the optimal zero-error
information transmission rate of a quantum channel. The
implication of the additivity conjecture is that entangled signal
states are not more useful than separable signal states for
communication through the quantum channel.

A related additivity conjecture is concerns the so-called
entanglement of formation $E_f$, which asymptotic behavior,
$\lim_{n\to\infty}\frac{1}{n}E_f(\rho^{\otimes n})$, named the
entanglement cost $E_c(\rho)$, is the number of maximally
entangled pairs required to prepare $\rho$ by LOCC in an
asymptotic way. The additivity of $E_f(\rho)$  implies that
$E_c(\rho)=E_f(\rho)$, simplifying the computation of $E_c(\rho)$
to large extent, and also that
$E_c(\rho\otimes\sigma)=E_c(\rho)+E_c(\sigma)$, meaning that
making $\rho$ and $\sigma$ altogether requires the same amount of
maximally entangled states as they are produced separately (in
other words, there is no catalytic effect in entanglement
 dilution, different from entanglement distillation).

There is yet another stronger conjecture about $E_f$, its
\textit{strong super-additivity} \cite{VW}. If $\rho$ is a state on
${\cal A}\otimes{\cal B}$, where ${\cal A}={\cal A}_1\otimes {\cal
A}_2$ and ${\cal B}={\cal B}_1\otimes {\cal B}_2$, then the strong
super-additivity  of $E_f$ is defined as the property~:
\be E_f(\rho)\geq E_f(\rho_1)+E_f(\rho_2),\ee
where all $E_f$ are taken with respect to the $\{A,B\}$ partition,
and $\rho_i$ are reduced density matrix on ${\cal A}_i\otimes{\cal
B}_i$. In short, the sum of local entanglements are smaller than
global entanglement.

As is pointed out in \cite{MSW}, the strong super-additivity
conjecture implies both of previously mentioned two
additivity conjectures. Peter Shor also proved that these
three additivity conjectures are indeed equivalent \cite{Shor}.
Hence, one of the proof of these conjectures will solve all the
problems at once. So far, despite many efforts, these additivity
conjectures have been proven only for some special instances \cite{Hiroshima,King1,King2,MSW,MY,Shor2,VDC,WolfEisert,VW}. In
this letter, we prove the strong super-additivity for a discrete
but diverse subclass of states, known as stabilizer states.

\subsection{Pure stabilizer states}
%
%
%

In a previous paper \cite{fattal04}, we have derived a very simple
expression for the entropy of entanglement of a stabilizer state of
$n$ qubits for a given bi-partition. Here, we use this result to
prove the strong super-additivity of the entanglement of
formation for stabilizer states.\\


Let $\Pa{n}$ denote the Pauli group for $n$ qubits. A pure
stabilizer states $\p$ of n qubits is a simultaneous eigenvector of
n independent pauli operators with eigenvalue $\pm 1$. The $n$
independent pauli operators generate an abelian subgroup of
$\Pa{n}$, called the stabilizer group of $\p$ et denoted $\S$.

Let $(A,B)$ be a partition of the $n$ qubits. Define $S_A$ (resp.
$S_B$) to be the subgroup of $\S$ containing operators that act
trivially on $B$ (resp. $A$)~:
\bea S_A &=& \{{g_A \otimes I_B \in
\S}\}\\
S_B &=& \{{I_A \otimes g_B \in \S}\} \eea
We will refer to $S_{loc} = S_A \cdot S_B$ as the local subgroup of
$S$. Then $\p$ is separable for partition $\{A,B\}$ if and only if
$S = S_{loc}$. Otherwise we can find a non-trivial subgroup $S_{AB}$
of $S$ such that~:
\be S = S_A \cdot S_B \cdot S_{AB} \label{S_decomp}\ee
$S_{AB}$, unlike $S_{loc}$, is not uniquely determined. If $S_{AB}$
satisfies (\ref{S_decomp}), and $g_{loc}$ is an element of
$S_{loc}$, then $g_{loc}\cdot S_{AB}$ also satisfies
(\ref{S_decomp}). The rank $e_{AB}$ of $S_{AB}$ - defined as its
minimum number of
generators - turns out to be twice the entropy of entanglement of $\p$.\\


We now introduce a different partition (1,2) of the same set of
qubits. We will exhibit a pure state statistical decomposition of
$Tr_2(\proj{\psi})$ in terms of stabilizer states sharing a common
stabilizer group up to phases. For this we will use $n_{A2}$
commuting independent pauli operators $M^{A2}_i$ with support on
$A2$, and $n_{B2}$ commuting independent pauli operators $M^{B2}_j$
with support on $B2$. Together these operators form a complete
commuting independent set for partition $2$, that is moreover local
for partition $\{A,B\}$. Measuring the commuting operators
$M^{A2}_i$'s on the qubits of $A2$ has $2^{nA2}$ possible outcomes
denoted by the binary string $k^{A2}$ of length $n_{A2}$, that we
can also use to label the mutually orthogonal post-measurement
states $\ket{k^{A2}}$. We define a similar basis with similar
notations for $B2$. It is then straightforward to see that an
unreferred successive measurement of the $M^{A2}$'s and $M^{B2}$'s
on the initial state $\p$ yields the mixed state~:
\bea \rho^M & =  \sum_{k^{A2},k^{B2}} &
\iprod{k^{A2},k^{B2}}{\psi}\iprod{\psi}{k^{A2},k^{B2}}\nonumber\\
& & \otimes \proj{k^{A2},k^{B2}} \eea
This state is a statistical mixture of stabilizer pure states on the
sets of all qubits, with common stabilizer group $S^M$ - up to
phases - given by~:
\be S^M = S^M_1 \cdot \<M^{A2}_i\> \cdot \<M^{B2}_j\> \ee
and obtained from $\S$ after the successive measurements by the
prescription described e.g. in \cite{Ike_book}.

Note that $\iprod{k^{A2},k^{B2}}{\psi}$ is a pure stabilizer state
on partition $1$, with stabilizer group $S^M_1$. We will write this
state $\psi(k^{A2},k^{B2})$ and rewrite the decomposition of
$\rho^M$ explicitly as~:
\be \rho^M = \sum_{k^{A2},k^{B2}}
\proj{\psi(k^{A2},k^{B2}),k^{A2},k^{B2}} \ee
The critical observation now is that the entropy of entanglement of
$\ket{\psi(k^{A2},k^{B2}),k^{A2},k^{B2}}$ with respect to partition
$\{A,B\}$ is the same as the entropy of entanglement of
$\ket{\psi(k^{A2},k^{B2})}$.\\

On the other hand, using the $\ket{k^{A2},k^{B2}}$ basis to take the
partial trace over partition 2, we get~:
\be Tr_2(\proj{\psi}) = \sum_{k^{A2},k^{B2}}
\proj{\psi(k^{A2},k^{B2})} \ee
which is also a decomposition into pure stabilizer states of same
stabilizer group $S^M_1$. Since the entanglement of formation of
$Tr_2(\proj{\psi})$ is defined as the minimum average entropy of
entanglement over all decomposition of $Tr_2(\proj{\psi})$ into pure
states, we obtain the key result~:
\be E_f(Tr_2(\proj{\psi})) \leq e_{AB}(S^M_1) \ee
A similar inequality hold for the partial trace over the partition
$1$.

We will now see that there is a systematic way to choose the
$M^{A2}$ and $M^{B2}$ operators, and similarly the $M^{A1}$ and
$M^{B1}$ operators such that~:
\be e_{AB}(S^M_1)+e_{AB}(S^M_2) \leq e_{AB}(S) \label{result}\ee
which will imply the strong super-additivity.\\

We start from the decomposition (\ref{S_decomp}) and further break
$S_A$ and $S_B$ as follows~:
\bea S_{A} &= & S_{A}^1 \cdot S_{A}^2 \cdot S_{A}^{12}\\
 S_{B} &= & S_{B}^1 \cdot S_{B}^2 \cdot S_{B}^{12}\eea
where $S_{A}^1$ is the subgroup of $S_A$ that acts trivially on the
qubits of partition 2.

Before we choose our measurement operators, we need a few lemmas to
guide our choice.


First we observe that the post-measurement state obtained after
measuring a local operator with respect to $\{A,B\}$ cannot have
more entanglement than the pre-measurement state. Simply the local
subgroup of the post-measurement state will either be of same size
or grow after the measurement, which means that the entanglement
stays the same or decrease. This is trivial if $M$ commutes with
$S_A$ and $S_B$, since the local subgroup after measurement will
contain $S_A \cdot S_B$ therefore at least stays of same size. If
$M$ does not commute with $S_A \cdot S_B$, then we can write~:
\be S_A \cdot S_B = <s_1,...,s_{n-e_{AB}}> \ee
such that $M$ anti-commutes with $s_1$ but commutes with all other
$s_j$ ($2\leq j \leq n-e_{AB}$) \cite{fattal04}. Then recalling that
$M$ is a local operator (i.e. acts on A only or B only), the local
subgroup after measurement will contain the subgroup
$<M,s_2,...,s_{n-e_{AB}}>$,
therefore will again be at least of same size as the pre-measurement local subgroup.\\

We now derive our most useful tool~:\\

\textbf{Lemma :} Measuring a local operator $M$ (acting trivially on
A or B) that commutes with $S_A$ and $S_B$ but that is not in the
stabilizer group S reduces the entanglement $e_{AB}$ of the
post-measurement stabilizer $S^M$ by at least 2.

\textit{proof :} $S_A$ and $S_B$ are preserved by the measurement,
but we also know that $M$ must lie in the disentangled subspace of
$S^M$, therefore the local subgroup must be at least contain $S_A
\cdot S_B \cdot \<M\>$. Therefore $e_{AB} = n-|S_{loc}|$ must
decrease by at least one unit. But we also know from \cite{fattal04}
that $e_{AB}$ must be an even number, hence we conclude it must have decreased by at least 2. $\Box$\\

Denote by $P_2$ the projection on partition 2, defined by~:
\be P_2 (g_1 \otimes g_2) \equiv I_1 \otimes g_2 \ee

We claim that $P_2(S_A^{12})$ is a subgroup of same rank as
$S_A^{12}$. Indeed, if $S_A^{12}$ is generated by some $\{g_k\}$,
then~:
\bea & \prod P_2(g_k) = I & \\
\Rightarrow & P_2(\prod g_k) = I \\
\Rightarrow & \prod g_k \in S_1 \cup S_A^{12} &=\{I\} \eea
and hence the $P_2(g_k)$ are independent. Also, the $P_2(g_k)$ must
all commute with $S_A^2$, since the $g_k$ do.

We can organize the generators of $P_2(S_A^{12})$ as follows~:
\be P_2(S_A^{12}) = Z \cdot <g_{j}, \bar{g}_{j}>_{j=1..p} \ee
where $Z$ is the center of $P_2(S_A^{12})$ (i.e. the subgroup that
commutes with all elements of the group), and the $(g_{j},
\bar{g}_j)$ form anti-commuting pairs, commuting with all other
generators. We can always find such generators up to some
multiplications by elements of the local subgroups, as shown in \cite{fattal04}.\\

For our measurement operators on $A2$, we pick the $\bar{g}_{j}$
first, that is choose $M^{A2}_i = \bar{g}_{i}$ for $1\leq i \leq p$,
$p$ being the number of non-commuting pairs in $S_A^{12}$. Note
that~:
\be p \leq \frac{1}{2}|S_A^{12}| \label{ineq}\ee
After the measurement of all $\bar{g}_{j}$, all the anti-commuting
pairs in $P_2(S_A^{12})$ are "destroyed", to the benefit of $S_A^1$
and $S_A^2$ which each gain a new independent element~: $S_A^2$
receives $\bar{g}_{j}$, and $S_A^1$ receives $\tilde{g}_{j}$ such
that $\tilde{g}_{j} \otimes \bar{g}_{j}$ was in the initial subgroup
$S_A^{12}$.

We then choose the remaining $M^{A2}_i$ ($i=p+1..n_{A2}$) to be the
generators of $S_A^2$, the generators of $Z$, and possibly complete
the list with elements that commute with $S_A$, $Z$, and the
$\bar{g}_{j}$.

Note that all these operators are local with respect to $\{A,B\}$.
Moreover, apart from the elements of $S_A^2$ and the $\bar{g}_{j}$,
they all satisfy the assumptions of the lemma, that is commute with
the local subgroup but are not element of the stabilizer group
itself. Measuring these operators adds one independent element to
both $S_A$ and $S_B$ and reduces the overall entanglement by $2$
each time. The number $N_{A2}$ of these operators is~:
\bea N_{A2} & = &  n_{A2}-|S_A^2|-p \\
& \geq & n_{A2}-|S_A^2|-\frac{1}{2}|S_A^{12}| \eea
using the inequality (\ref{ineq}). The rest of the operators we
measure on $B2$ are also local with respect to $\{A,B\}$, and
therefore do not increase the entanglement.

Now had we chosen to trace over $B2$ rather than $A2$ first, a
similar inequality would hold on subspace $B2$. Therefore the total
number $N$ of measurement operator that reduce the overall
entanglement (each by 2) is at least~:
\bea N & \geq & max(N_{A2}, N_{B2}) \\
& \geq &
\frac{1}{2}(n_2-|S_A^2|-|S_B^2|-\frac{1}{2}|S_A^{12}|-\frac{1}{2}|S_B^{12}|)
\eea
where we used the fact that the $max(N_{A2}, N_{B2}) \geq
\frac{N_{A2}+ N_{B2}}{2}$.

The residual entanglement $e_{AB}(S^M_1)$ common to the pure states
on partition 1 described by the stabilizer group $S^M_1$ resulting
from the measurement of operators $M^{A2}$ and $M^{B2}$ therefore
obeys the inequality~:
\bea  e_{AB}(S^M_1) &\leq& e_{AB}(S) - 2N\\
 & \leq & e_{AB} - n_2 +
 |S_A^2|+|S_B^2|+\frac{1}{2}|S_A^{12}|+\frac{1}{2}|S_B^{12}| \eea
Similarly, after partial trace over 1, we obtain a similar
inequality~:
\be e_{AB}(S^M_2) \leq e_{AB}(S) - n_1 +
|S_A^1|+|S_B^1|+\frac{1}{2}|S_A^{12}|+\frac{1}{2}|S_B^{12}| \ee
Finally using the fact that~:
\be |S_A^1|+|S_A^2|+|S_A^{12}| = |S_A| \ee
and that~:
\be |S_A|+|S_B|= n - e_{AB} \ee
we obtain the announced result (\ref{result}).\\

To summarize, we have proven that the (possibly mixed) state
obtained after partial trace over partition 2 (resp 1) had an
entanglement of formation that was upper bounded by the entropy of
entanglement corresponding to the stabilizer pure states defined on
the global set of qubits and obtained after measuring a complete set
of independent commuting operators on partition 2 which are local
for the $\{A,B\}$ partition. We could explicit a set of such
operators for both partitions 1 and 2 so that the sum of the entropy
of entanglement of the post-measurement stabilizer groups was
smaller than the original entropy of entanglement for the pure state
$\p$, which is equal to its entanglement of formation since it is a
pure state.


\subsection{Mixed states extension}
The result generalizes easily to mixed states for which the
decomposition into pure state that minimizes the average entropy of
entanglement happens to be a mixture of stabilizer states with same
stabilizer group - again up to phases.

Indeed, suppose that the $\psi_j$'s are stabilizer pure states
described by same stabilizer group $S$, and that~:
\be E_f(\rho = \sum_j p_j \proj{\psi_j}) = \sum_j p_j E(\psi_j) \ee
We can write the partial trace of $\rho$ over partition 2 as~:
\be Tr_2(\rho) = \sum_j p_j Tr_2(\proj{\psi_j}) \ee
Now $Tr_2(\proj{\psi_j})$ is a mixture of pure stabilizer states
with common stabilizer group $S^M_{1j}$ and $Tr_1(\proj{\psi_j})$ is
a mixture of pure stabilizer states with common stabilizer group
$S^M_{2j}$ which, according to the previous analysis, verifies~:
\be e_{AB}(S^M_{1j}) + e_{AB}(S^M_{2j}) \leq e_{AB}(S) \ee
Now since~:
\bea E_f(Tr_2(\rho)) &\leq & \sum_j p_j e_{AB}(S^M_{1j}) \\
E_f(Tr_1(\rho)) &\leq & \sum_j p_j e_{AB}(S^M_{2j}) \eea
we obtain the announced result~:
\be E_f(Tr_2(\rho)) + E_f(Tr_1(\rho)) \leq e_{AB}(S) \ee

We finally prove that mixed stabilizer states of the form~:
\be \rho = \sum_{g \in H} g \label{rhodec}\ee
where $H$ is a non-maximal abelian subgroup of $\Pa{n}$ have the
property mentioned above, namely minimize the average entropy of
entanglement when written as a mixture of pure stabilizer states
with common stabilizer group $S$ - up to phases.\\

To prove this result, we use the fact that $H$ - as any stabilizer
group - is LU-equivalent to a stabilizer state for a which a
stabilizer generator list is composed of single $Z$ operators acting
on single qubits of partition $A$ or $B$, single $ZZ$ operators
acting simultaneously on a qubit of $A$ and a qubit of $B$, and a
number $p$ of locally anti-commuting pairs $XX,ZZ$ also acting
across the $\{A,B\}$ partition but yet on a disjoint set of qubits
{\bf [ref]}. We will denote the corresponding local unitaries $U_A$
and $U_B$. We will prove that $E_f(\rho)$ is precisely $p$ the
number of locally anti-commuting pairs, and will exhibit a pure
stabilizer state decomposition of $\rho$ for which the average
entropy of entanglement is $p$. We will start by exhibiting such a
state. Consider the non-maximal stabilizer group $\tilde{H} = U_AU_B
H U_A^{\dagger}U_B^{\dagger}$ and its decomposition into single
$Z$'s, $ZZ$'s, and pairs $(XX,ZZ)$'s acting on disjoint supports. We
complete the stabilizer generator list of $\tilde{H}$ with single
$Z$ operators acting on whatever qubit lays out of the support of
the generators of $\tilde{H}$, and a single $Z$ operator acting on
the first qubit of every single $ZZ$ operator featuring in that
generator list. It is easy to check that the stabilizer group
$\tilde{S}$ obtained in this way is a maximal stabilizer group
having $\tilde{H}$ as a subgroup. Hence $\rho$ can be expressed as a
statistical mixture (with equal weights) of pure stabilizer states
all having stabilizer group
$U_A^{\dagger}U_B^{\dagger}\tilde{S}U_AU_B$, and therefore with
average entropy of entanglement equal to $p$. This proves that
$E_f(\rho) \leq p$.

Now suppose that we can decompose $\rho$ as~:
\be \rho = \sum_j p_j \proj{\psi_j} \ee
where the $\ket{\psi_j}$'s are arbitrary pure states. From
(\ref{rhodec}), we see that for any $g \in H$, $g \rho = \rho$ which
after the trace implies~:
\be \sum_j p_j \bra{\psi_j}g\ket{\psi_j} = 1 \ee
which itself implies that $\bra{\psi_j}g\ket{\psi_j} = 1$ for all
$j$, and that~:
\be g\ket{\psi_j} = \ket{\psi_j} \label{eigeq} \ee
We will consider for $g$ the first pair of locally anti-commuting
generators that we write $(g_Z,g_X)$, for which we can find suitable
local unitaries $U_A, U_B$ such that~:
\bea g_X & = & U_AU_B X_{A1}X_{B1} U_A^{\dagger}U_B^{\dagger}\\
g_Z & = & U_AU_B Z_{A1}Z_{B1} U_A^{\dagger}U_B^{\dagger}\eea
where $A1$ and $B1$ designate the first qubit of partition $A$ and
$B$ respectively. It is then not hard to see that due to
(\ref{eigeq}), the pure state $U_A U_B\ket{\psi_j}$ separate as a
Bell pair on qubits $A1,B1$, and a pure state on the remaining
qubits. Iterating this argument, we see that each state $U_A
U_B\ket{\psi_j}$ contains at least $p$ Bell pairs on disjoint qubit
pairs, and therefore has entropy of entanglement at least $p$. This
proves that $E_f(\rho) \geq p$, and therefore that the stabilizer
construction above yields the decomposition of $\rho$ into a mixture
of pure states that achieves the minimum average entropy of
entanglement. 






\end{document}